\documentclass[fleqn,10pt]{wlscirep}
\title{Capsize of polarization in dilute  photonic crystals}
\author[1,2*]{Zhyrair Gevorkian}
\author[2]{Arsen Hakhoumian}
\author[3]{Vladimir Gasparian}
\author[4]{Emilio Cuevas}
\affil[1]{Yerevan Physics Institute, Alikhanian Brothers St. 2, 0036 Yerevan, Armenia}
\affil[2]{Institute of Radiophysics and Electronics, Ashtarak-2, 0203, Armenia}
\affil[3]{California State University, Bakersfield, California 93311-1022, USA}
\affil[4]{Departamento de F\'{i}sica, Universidad de Murcia, E-30071 Murcia, Spain}
\affil[*]{gevork@yerphi.am}
\begin{abstract}
We investigate, experimentally and theoretically, polarization rotation effects in dilute photonic crystals with transverse permittivity inhomogeneity perpendicular to the traveling direction of waves. A capsize, namely a drastic change of polarization to the perpendicular direction is observed in a one-dimensional photonic crystal in the frequency range $10\div 140$ GHz. To gain more insights into the rotational mechanism, we  have developed a theoretical model of dilute photonic crystal, based on Maxwell's equations with a spatially dependent two dimensional inhomogeneous dielectric permittivity. We show that the polarization's rotation can be explained by an optical splitting parameter appearing naturally in Maxwell's equations for magnetic or electric fields components. This parameter is an optical analogous of Rashba like spin-orbit interaction parameter present in quantum waves, introduces a correction to the band structure of the two-dimensional Bloch states,  creates the dynamical phase shift between the waves propagating in the orthogonal directions and finally leads to capsizing of the initial polarization.
Excellent agreement between theory and experiment is found.
\end{abstract}
\begin{document}

\flushbottom
\maketitle

\thispagestyle{empty}








\section{Introduction}

The problem of the polarization rotation (PR) of an incident electromagnetic wave is one of the most interesting in photonic crystals (PC)\cite{Joan08}. The investigations in this field are stimulated by the possibility of creation of polarization controlling devices, where the polarization of the propagating light could be precisely manipulated and controlled\cite{solli2004,li2001,solli2003}. One of the mechanism that can account for this PR can be associated with the geometrical anisotropy of photonic crystal (PC)\cite{shaini}. The anisotropy,  imposing asymmetry into a periodic two dimensional (2d) structure in $xy-$plane, leads to a different speed in the $x-$ and $y-$axes resulting in a phase delay between $H_x$ ($E_x$) and $H_y$ ($E_y$) components. Another mechanism that can lead to the PR of propagating light is due to dielectric permittivity inhomogeneity of medium. This phenomenon known long ago \cite{Rytov,Vlad,Kravor80,Vinit90} and accepted view of the physics behind the effect is that the rotation is controlled by the parameter $\lambda/\Lambda_0$ ($\lambda$ is the wave length and $\Lambda_0$ is the typical length scale of the system's inhomogeneity). In the geometrical optics approximation when wavelength is much smaller than inhomogeneity characteristic scale, rotation angle is negligible \cite{Bliokh03}. However, in the opposite limit, the rotation of polarization due to dielectric permittivity inhomogeneity, is quite significant. Particularly, such effects are observed both in dielectric (see Ref. \cite{dielpolrot12} and references therein) and metallic\cite{metalphoton} PCs.

In this paper we investigate, experimentally and theoretically, the PR effects in a dilute photonic crystals (DPC) with transverse permittivity inhomogeneity perpendicular to the $z-$traveling direction of waves.
Interest to DPC is largely motivated by a peculiar behavior of the transmission coefficient, recently studied in Ref. \cite{GGC16}. Using parabolic (forward) scattering approximation to investigate electromagnetic wave propagation in inhomogeneous media \cite{FF79,Lag}, in Ref. \cite{GGC16} was found an independence of the transmission coefficient of the central diffracted wave from the incident wavelength in a dilute perforated metal system.  As we will see below, it turns out that the DPC along with the interesting transport properties lead to unusual polarization effects as well.

Our direct measurements indicate that under certain conditions the initial polarization changes its direction by $\pi/4$ in 1d DPCs in the frequency range of $10\div 140$ GHz (see below).
The main experimental finding consists of not typical shift of distribution's minimum and maximum regions in the spectra of the transmitted intensity (see below). To understand and explain the rotating mechanism of the DPC, we  have developed a theoretical model, based on Maxwell's equations with a spatially dependent two dimensional dielectric permittivity $\varepsilon(x,y)$. Using the same parabolic approximation, mentioned above, we show that the polarization's rotation in 1d and 2d DPCs can be explained by an optical splitting parameter appearing naturally in Maxwell's equations for magnetic or electric fields components. This term is an optical analogous of Rashba type spin-orbit interaction parameter, introduces a correction to the band structure of the two-dimensional Bloch states,  creates the dynamical phase shift between the waves propagating in the orthogonal directions and finally leads to capsizing of the initial polarization. A photonic analog of Rashba type term in homogeneous chiral medium was considered in \cite{yanopapas11}.

Note that one should differ polarization rotation due to inhomogeneity and chirality or magnetoactivity of the medium. In the last two cases the dielectric permittivity tensor has non-diagonal terms in contrary to the former case. Non-diagonal terms lead to different velocities of left hand and right hand circular polarized waves in a chiral or magnetoactive medium. In PC situation is different. Here two waves polarized in orthogonal directions have different velocities. This difference leads to polarization rotation because every linearly polarized wave can be presented as a superposition of waves in two orthogonal directions. The possibility of large polarization rotation by PC was noticed in \cite{modopt95}.
\section{Experiment}

To study capsize effect in 1d PC system, we have performed an experiment where electromagnetic waves propagate through the dilute structure. The schematic setup of the experiment is shown in Fig. 1.
\begin{figure}
 \begin{center}
\includegraphics[width=8.0cm]{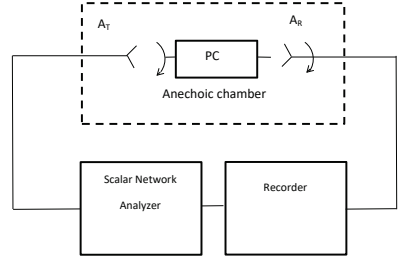}
\caption{Schematic of the experimental setup. Waves propagate through the dilute PC structure.
The arrangement is used to detect the frequency dependence of the normalized intensity $I$.}
\label{fig.1}
\end{center}
\end{figure}
It consists of transmitter and receiver antennas, an anechoic chamber with photonic crystal, scalar network analyzer and recorder. One-dimensional photonic crystal is formed from $7$ alumina ceramic plates with permittivity $\varepsilon=10$, thickness $b=0.5$ mm and sizes $60\times 96$ mm$^2$. They are parallel substituted along the $x-$axis at the distance $a=10$ mm from each other. The structure is supported by the low loss and low dielectric permittivity $\varepsilon=1.07$ foam layers, see Fig. 2.

 \begin{figure}
 \begin{center}
\includegraphics[width=8.0cm]{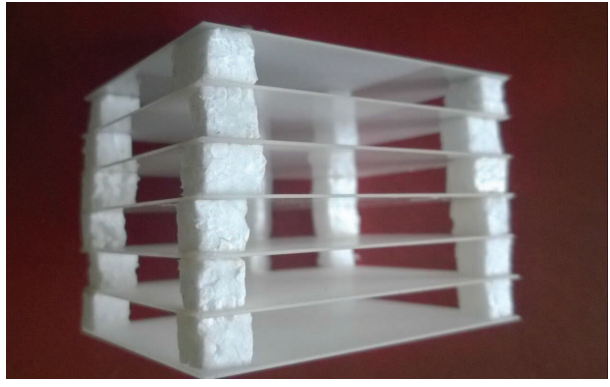}
\caption{One dimensional photonic crystal. Photograph of the 1d PC used in the experiment.
It is formed from $7$ alumina ceramic plates with $\varepsilon=10$, thickness $b=0.5$ mm, sizes
$60\times 96$ mm$^2$ and $a=10$ mm distance from each other.}
\label{fig.2}
\end{center}
\end{figure}
The measurements were carried out in the frequency range $10\div 140$ GHz (wavenumber $k_0=0.21\div 2.93$ mm$^{-1})$ using set of scalar network analyzers. Open end rectangular waveguides of corresponding frequency range are used as a transmitting and receiving antennas. Such a choice of antennas allow high level cross-polarization isolation smaller than $10^{-3}$
in their mutual orthogonal location. The PC and the receiving antenna can be rotated in the $xy-$plane, where the electric field vector ${\bf E}(\bf H)$ is located. We remind that the wave incidence direction is $z-$axis and plates are located in the $yz$-plane. The PC and the antennas were placed in an anechoic chamber to avoid the influence of unwanted external signals. The distances between them were chosen large enough that far field approximation is applicable. The frequency dependence of transmission coefficient was displayed on the recorder.
The results of experiment with one dimensional photonic crystal are presented in Fig. 3. In the upper panel (black line) the frequency dependence of transmission coefficient $I$ (normalized intensity) is shown provided that wave's incidence direction ($x-$axis) is perpendicular to the plates. Due to the periodicity in $x-$direction we expect to have frequency bands over which the propagation of the waves is forbidden, that is the transmission coefficient becomes zero. The panel indicates that at wave numbers above 0.48 mm$^{-1}$ the forbidden band begins to form.
In the middle and lower panels (red and blue lines, respectively) the wave incidence direction is $z-$axis, as discussed below. However, the plots differ from each other by the direction of receiving antenna: in the middle panel it is orthogonal to transmitter antenna ($y-$axis) and in the lower panel it is parallel to transmitter, that is directed in $x-$axis. The maxima in the middle panel corresponds to the resonance wavelengths $v_{1d}L=\pi/2+\pi n$(see theory below) for which polarization change the direction from $\pi/4$ (initial linear polarization angle) to $\pi/2$. In the lower panel (blue line) for the same wavelengths one has minimums instead of maximums. This shift of the intensity distribution is due to the change of receiver's direction: the latter now is pointed out on  $x-$direction, that is parallel to the transmitter polarization.  To verify the preceding discussion on the PR in 1d PC around resonance wavelengths $v_{1d}L=\pi/2+\pi n$, one needs to evaluate $v_{1d}$. Using equation  (\ref{spl1d}) we have calculated $v_{1d}(k_0)$ for the resonant wavelength $k_0=0.49$mm$^{-1}$ and for $\cos v_{1d}(k_0=0.49)L$ we get $-0.008$ ($L=96mm$) which is very close to $0$, as was predicted. Note the logarithmic scale used in the experimental plot Fig. 3. This means that the capsize phenomena takes place with very high accuracy $10^{-3}$.

\begin{figure}
\begin{center}
\includegraphics[width=10.0cm]{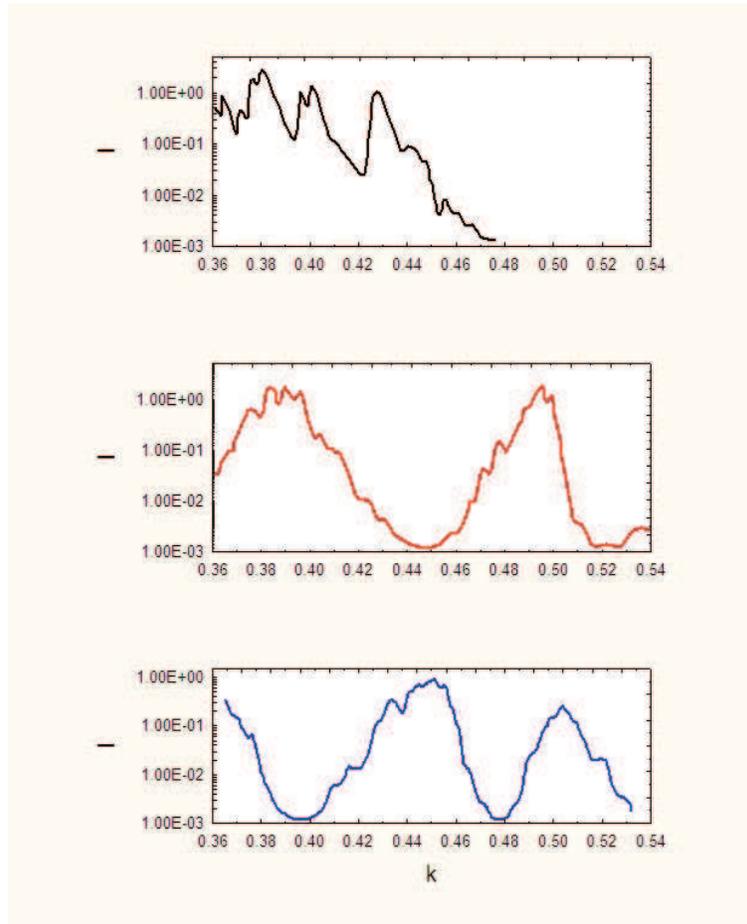}
\caption{Measured intensity for different wave's incidences.
The upper panel (black line) shows intensity when the wave normally falls on the parallel
plates that make one-dimensional photonic crystal. It indicates that at wave numbers above 0.48 mm$^{-1}$ the forbidden band begins to form in the frequencies range of interest. In the middle and lower panels (red and blue lines, respectively) the wavevector is parallel
to the plates.  However, the receiver is pointed out in $y-$direction (red line) and in
$x-$direction (blue line). The incident wavevector is on $z-$axis and the photonic crystal
is rotated on $\pi/4$ in the $xy-$plane -therefore initial polarization is at $\pi/4$.}
\label{fig.3}
\end{center}
\end{figure}

As we mentioned above, at $\theta_0=\pi/4$ and for resonant frequencies $v_{1d}L=\pi/2+\pi n$, the transmitted wave is linearly polarized but directed on $y-$axis. This means that the theoretical imaginary part of rotation angle that describes the ellipsity (ratio of ellipse axes) should become infinite.  However, in general, for arbitrary $\theta_0$ there is a finite ellipticity and the imaginary portion ${\rm Im} \theta(L)$ is also finite. In Fig. 4 we compare experimentally measured ellipticity at resonance wavenumber $k_0=0.49$ mm$^{-1}$ (red line) with the theoretical ellipticity (blue line) (see below, equation  (\ref{im1d})).
\begin{figure}
 \begin{center}
\includegraphics[width=8.0cm]{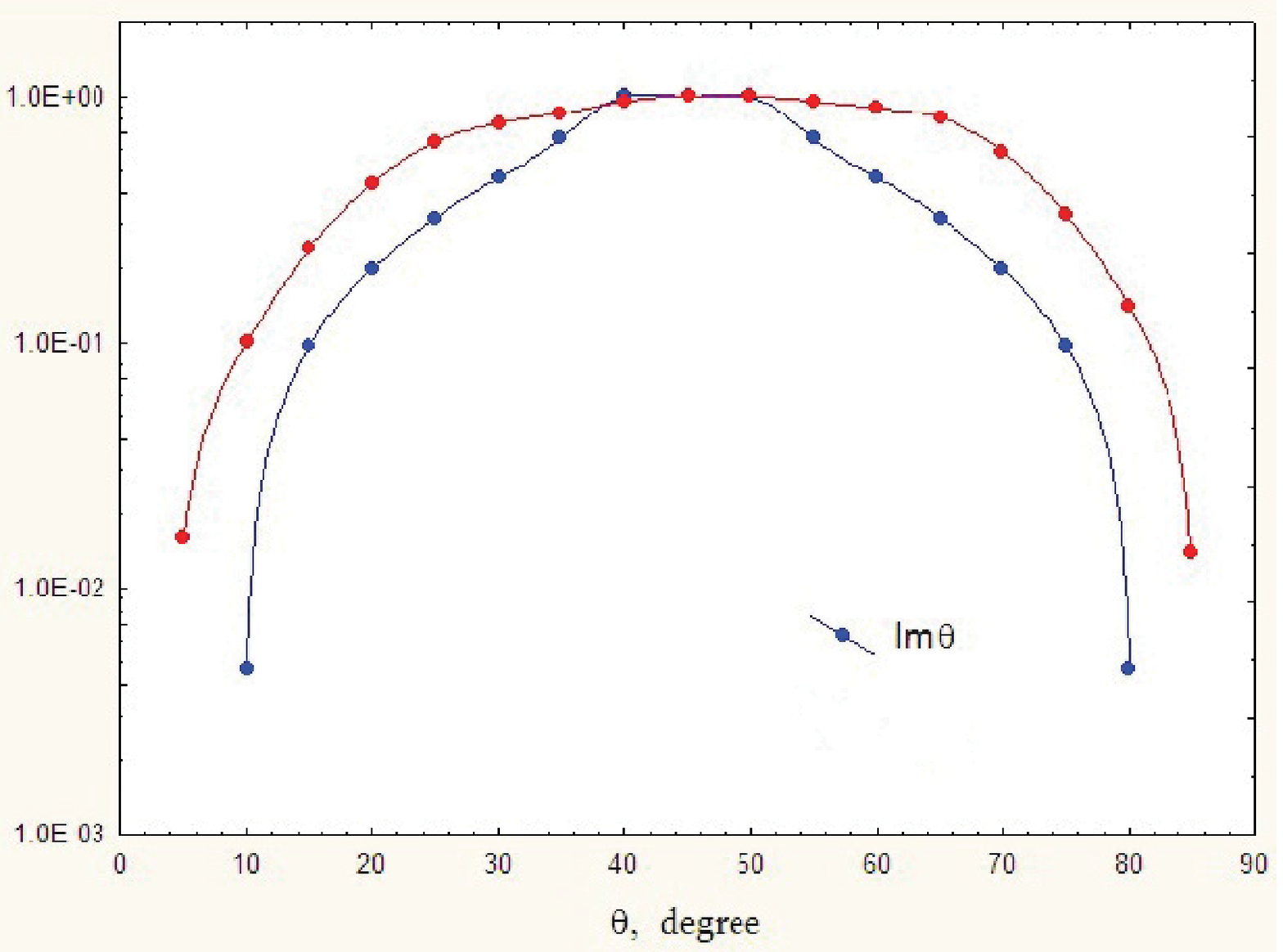}
\caption{Ellipticity. This plot shows the ellipticity dependence on initial linear polarization angle:
direct comparison betweeen experimentally measured ellipticity at resonance wavenumber $k_0=0.49$ mm$^{-1}$
(red symbols) with the theoretical ellipticity (blue symbols) given by equation  (\ref{im1d}).}
\label{fig.4}
\end{center}
\end{figure}

\section{Theory}

In achieving a better understanding of the experimentally observed capsize effect, discussed in the previous section, we develop an analytical approach that predicts the capsize of polarization in 1d and 2d DPCs.

Let us consider a dielectric medium with periodic array of two dimensional holes.
The inhomogeneity of the medium in $xy-$plain will be taken into account through a spatially dependent two dimensional dielectric permittivity $\varepsilon(x,y)$. Suppose a plane wave enters the medium from the $z<0$ half-space at normal incidence. Using Maxwell equations one can obtain a wave equation for $\bf H$ in the form
\begin{eqnarray}\label{sqeq}
\nabla^2{\bf H}(x,y,z)+k_0^2\varepsilon(x,y){\bf H}(x,y,z)+\\
\nonumber+\frac{\nabla\varepsilon(x,y)}{\varepsilon}\times{\bf \nabla}\times {\bf H}(x,y,z)=0,
\end{eqnarray}
where $k_0=\omega/c$ is the wave number corresponding to the angular frequency $\omega$ of the incident wave.
The last term in l.h.s. of equation (\ref{sqeq}) accounts for the contribution of the dielectric permittivity inhomogeneity and, as we will see below, introduces a correction to the band structure of the two-dimensional Bloch states. This correction creates the dynamical phase shift between the waves propagating in the orthogonal directions in the $xy-$plain and finally leads to capsizing of the initial polarization of $\bf H$.

Before proceeding further, we would like to mention that similarly to equation  (\ref{sqeq}) for $ \bf H$, one can write a wave equation for the electric field $\bf E$. However, the wave equation for $\bf E$ will contain en extra term ${\bf E}\nabla({\bf \nabla}\varepsilon/ \varepsilon)$ that does not affect the initial polarization. For this reason and for simplicity in what follows, we study only the peculiar rotation of the $\bf H$ in $xy-$plain caused for the dielectric permittivity inhomogeneity.

Following Refs. \cite{FF79,Lag,GGC16}, we seek the solution for the propagation in the system wave as a product of a fast and a slowly varying, $\bf\tilde{H}(x,y,z)$, function on a wave incident $z-$direction
\begin{equation}
{\bf H}(x,y,z)=e^{ik_0z} {\bf \tilde{H}}(x,y,z).
\label{slfas}
\end{equation}
Substituting equation (\ref{slfas}) into equation (\ref{sqeq}) and neglecting the second derivative of $\tilde{H}$ with respect to $z$ one gets in the parabolic equation approximation
\begin{equation}
i\frac{d\tilde{{\bf H}}}{dz}=\hat{H}(x,y)\tilde{{\bf H}},
\label{sreq1}
\end{equation}
where $\tilde{{\bf H}}\equiv\binom{\tilde{H}_x}{\tilde{H}_y}$ is a spinor and $\hat{H}=\hat{H}_0+\hat{V}$ is a $2\times 2$ matrix.

The explicit form of
$\hat{H}_0$ is
\begin{equation}
  \hat{H}_0=
\begin{pmatrix}
 -\frac{1}{2k_0}\nabla_\bot^2+\frac{k_0}{2}(1-\varepsilon(x,y)) & 0 \\
 0 & -\frac{1}{2k_0}\nabla_\bot^2+\frac{k_0}{2}(1-\varepsilon(x,y))  \label{tem}
\end{pmatrix}
\end{equation}
where $\nabla_\bot^2\equiv \partial^2/\partial x^2+\partial^2/\partial y^2$
and the inhomogeneity term is given by
\begin{equation}\label{inhom}
 \hat{V}=\frac{1}{2k_0\varepsilon}
\begin{pmatrix}
-\frac{\partial \varepsilon}{\partial y}\frac{\partial}{\partial y} & \frac{\partial \varepsilon}{\partial y}\frac{\partial}{\partial x}\\
\frac{\partial \varepsilon}{\partial x}\frac{\partial}{\partial y} & -\frac{\partial \varepsilon}{\partial x}\frac{\partial}{\partial x}
\end{pmatrix}.
\end{equation}
Let us stress that ignoring the second
derivative of $\tilde{H}$ with respect to $z$ in equation  (\ref{sreq1}) is justified by the fact that in a dilute system with a small fraction of dielectric (or metal) the light propagates in $z-$direction mostly forwardly, without back scattering.

So the problem of propagation of electromagnetic waves in a dilute dielectric medium with periodic array of two dimensional holes is reduced to a quantum motion of a particle with mass $k_0$ in a $2d$ periodic potential $V(x,y)=\frac{k_0}{2}\left(1-\varepsilon(x,y)\right)$ with inhomogeneity term $\hat{V}$, where spatial coordinate $z$ plays the role of time in a Schr{\"o}dinger-like
equation. In this approach, if the initial parameters at $z=0$ are known (for example the incoming wave front), then one can find them at $z>0$.  Such a mapping has been performed previously \cite{Lag,GGC16} and made it possible to study the propagation of light through a semi-infinite medium with transverse disorder \cite{Lag} and through  a dilute metal with two-dimensional hole arrays \cite{GGC16}.

It can furthermore be shown that if one associates $\tilde{H}_{x,y}$ components with spin components then the term $\hat{V}$ reminds Rashba spin-orbit interaction (see for a review Ref. \cite{Rashba15}, for photon spin-orbit interaction  \cite{Libzel92} and the review\cite{Bliokh15}). The remarkable similarity between $\hat{V}$, equation  (\ref{inhom}), and a Rashba spin-orbital interaction term
allow us to stress that both are closely related phenomena. The term $\hat{V}$ will introduce a correction to the band structure of the two-dimensional Bloch states, creates the dynamical phase shift between the waves propagating in the $x-$ and $y-$directions and finally leads to rotation of the initial polarization.
Hence, the calculation of the spectrum of Hamiltonian $\hat{H}(x,y)=\hat{H}_0+\hat{V}$ could provide a sensitive means to analyze polarization effects in dilute photonic crystal and may be used as a starting point to evaluate the wave transmission coefficient at $z-$direction.

To find the band structure let us represent the solution of equation (\ref{sreq1}) through the eigenfunctions of the Hamiltonian equations  (\ref{tem}) and (\ref{inhom}),
\begin{equation}
\tilde{{\bf H}}(x,y,z)=\sum_nc_ne^{-iE_nz}{\bf h}_n,
\label{eig}
\end{equation}
and
\begin{equation}
\hat{H}{\bf h}_n(x,y)=E_n{\bf h}_n(x,y).
\label{eigfun}
\end{equation}

Finally, substitution of equation (\ref{eig}) into equation (\ref{slfas}) yields the solution of Maxwell equation
\begin{equation}
{\bf H}(x,y,z)=e^{ik_0z}\sum_nc_ne^{-iE_nz}{\bf h}_n(x,y),
\label{maxsol}
\end{equation}
which can serve as a basis for qualitative understanding of spectral properties of dilute systems. Such an approach allows one to pinpoint the specific details and provide information
about the energy band structure.

To close this section let us note that if the dielectric
permittivity $\epsilon (x, y)$ is a periodical function, then the spectrum
of the Hamiltonian (\ref{tem}) consists of allowed
and forbidden energy bands, if the term $\hat{V}$ is zero. If  $\hat{V}\neq 0$ then the electronic energy bands are split by $\hat{V}$ and as a consequence the polarization of the propagated light is rotated.

\section{Spectrum of photonic crystal}

Before discussing the physics induced by $\hat{V}$ coupling let us consider $\hat{V}=0$. For this case the quantum-mechanical problem
is reduced to the well-known Kronig-Penney (KP) model due to the fact that the periodic $\varepsilon(x,y)$ spectrum of Hamiltonian consists of energy bands separated by gaps and wave functions described by Bloch states.
In this section we will investigate band structure for symmetric KP model in one and two dimensional systems for
quasimomentum $q=0$, that is at the center of Brillouin zone. Note that only these states give contribution to the
transmitted light intensity in the normal direction\cite{GGC16}.

We first discuss a two-dimensional symmetric KP model where the dielectric function has the form
\begin{eqnarray}
  \varepsilon(x,y) =\left\{
\begin{array}{rl}
1 & \mbox{if $0<x,y<a-b$} \\
\varepsilon & \mbox{if $a-b<x$ or $a-b<y$}
\end{array} \right. \label{ep2}
\end{eqnarray}
The Bloch state wave function $u(x,y)\equiv u_{q=0}(x,y)$, using periodicity of $u(x,y)$ and its derivatives,  can be represented in the form that are valid for the energy $E_0\equiv E(q=0)$ that must satisfy the dispersion relation (see below, Eq. (\ref{disp})).

\begin{eqnarray}
u_{1}(x,y)=\frac{A_2}{\cos\frac{\beta(a-b)}{2}}\cos\beta \bigg (x+y-\frac{a-b}{2}\bigg)\quad 0<x,y<a-b \nonumber \\
u_{2}(x,y)=\frac{A_2}{\cos\frac{\alpha b}{2}}\cos\alpha \bigg (x+y+
\frac{b}{2}\bigg),\quad a-b<x\quad or \quad a-b<y
\label{wavefun}
\end{eqnarray}
where $\beta=\sqrt{k_0E_0}$ and $\alpha=\sqrt{k_0(V_d+E_0)}$, $E_0>0$, $V_d=k_0(\varepsilon-1)/2$. The parameter $A_2$ can be found from the normalization condition $\int_{0}^{a}dxdy u^2_{q=0}(x,y)=1$.
The Bloch state wave function $u(x,y) \equiv u_{q=0}(x,y)$  obeys the Schr\"{o}dinger equation
\begin{equation}
\left[-\frac{1}{2k_0}\nabla_\bot^2+V(x,y)\right]u(x,y)=E_0u(x,y),
\label{shrod}
\end{equation}
 and energy $E_0\equiv E(q=0)$ (for $q=0$ state) is determined from the similar to the one dimensional Kronig-Penney dispersion relation
\begin{equation}
1=\cos\alpha b\cos\beta(a-b)-\frac{\alpha^2+\beta^2}{2\alpha\beta}\sin\alpha b\sin\beta(a-b).
\label{disp}
\end{equation}
Putting all of this information together one can, at least numerically, evaluate not only the energy $E_0$, using equation  (\ref{disp}), but also calculate the intensity of central diffracted wave in $2d$ system in terms of dielectric (metal) fraction.

As for the $1d$ PC case (dielectric permittivity $\varepsilon(x)$ periodically depends only one coordinate) it can be obtained from above formulas by the following substitutions (assuming that $y\equiv 0$): $\beta=\sqrt{2k_0E_0}$ and $\alpha=\sqrt{2k_0(V_d+E_0)}$.
We have presented the photonic band structure of transversal motion for the $1d$ crystal used in our experiment in Fig.5.
\begin{figure}
 \begin{center}
\includegraphics[width=8.0cm]{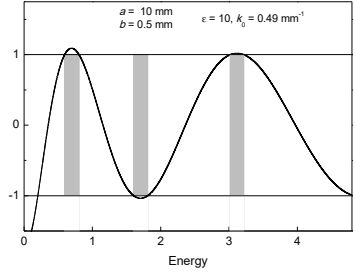}
\caption{Transversal photonic band structure. The vertical axis is the dispersion equation right side Eq.(\ref{disp}) calculated for the $1d$ crystal used  in our experiment Fig.2 at the resonance wavenumber $k_0=0.49mm^{-1}$, Fig.3. The shaded parts are the forbidden bands.}
\label{fig.5}
\end{center}
\end{figure}

The border value $E_0\equiv E(q=0)=0.401<k_0$ is the transversal wavenumber. The incident wavenumber $k_0$ lies in the forbidden band.
After a brief sketch of the steps of calculating of spectrum of periodic PC in $1d$ and $2d$ systems we start a detailed discussion of the interaction $\hat{V}$ role in a PR. We will see below that in some cases the interaction term $\hat{V}$  plays a central role of the capsizing of initial polarization.

\section{Polarization rotation}

\subsection{Two dimensional photonic crystal}

As mentioned above the interaction $\hat{V}$ splits $E_0$ into two energy levels. To find energy splitting in the
mean field approximation in a $2d$ case, we substitute the operator $\hat{V}$ by its expectation value
and diagonalize $2\times 2$ matrix Hamiltonian
\begin{equation}\label{split}
\hat{H}=
\begin{pmatrix}
E_0-v_{2d}& v_{2d}\\
v_{2d}   & E_0-v_{2d}
\end{pmatrix},
\end{equation}
where
\begin{equation}\label{2dspint}
v_{2d}=\frac{1}{2k_0}\int_{0}^{a}dxdy\frac{1}{\varepsilon(x,y)}\frac{\partial \varepsilon(x,y)}{\partial y}u(x,y)\frac{\partial u(x,y)}{\partial y}.
\end{equation}
Note that  for the symmetrical Kronig-Penney model all the integrals in $\hat{V}$ are equal to $v_{2d}$.
Direct calculation  of the integral (\ref{2dspint}) by using  (\ref{wavefun}) yields
\begin{equation}
v_{2d}=A_2^2\frac{\varepsilon-1}{2k_0\varepsilon}\tan{\frac{\alpha b}{2}}{\sin2\alpha b}
\label{split2d},
\end{equation}
where $A_2$, as was mentioned, is the normalization parameter with dimensionality $a^{-1}$ (a is the period of the system).

For the magnetic field of central diffracted wave we get
\begin{equation}\label{finalfield}
  \binom{H_x(x,y,z)}{H_y(x,y,z)}=\left [e^{i(k_0-E_1)z}{\bf h_1}+e^{i(k_0-E_2)z}{\bf h_2}\right ]u(x,y).
\end{equation}

One can interpret this result as an PR due to the interaction term $\hat{V}$ or Rashba splitting parameter $v$. The latter,  imposing asymmetry into a periodic 2d structure in $xy-$plane, leads to a different speed in the x and y axes resulting in a phase delay between $H_x$ and $H_y$ components.

Introducing parameter $c=c_1/c_2$, the PR angle
at $z=L$ can be written in the form
\begin{equation}\label{angle}
\tan\theta(L)=\frac{H_y(z=L)}{H_x(z=L)}=\frac{c^2-1-2ic\sin (E_1-E_2)L}{c^2+2c\cos (E_1-E_2)L+1},
\end{equation}
provided that an initial polarization angle $\theta_0$ at $L=0$ is given by
 \begin{equation}\label{inangle}
\tan \theta_0= \frac{c-1}{c+1}.
\end{equation}
By applying equation  (\ref{inangle}) and calculating the real part of rotation angle $\theta$ from equation  (\ref{angle})
one obtain ($E_1-E_2\equiv 2v_{2d}$)

\begin{equation}\label{finalangle}
{\rm Re}\, \theta(L)=\frac{1}{2} {\rm arccot} (\cot2\theta_0\cos2v_{2d}L),
\end{equation}
where the principal interval of the arccotangent function is chosen $0\le {\rm arccot}\, x <\pi$ to
avoid discontinuity at $x = 0$.

The above expression ${\rm Re}\, \theta(L)$ is a general expression for PR angle and valid for  dilute photonic crystal with inhomogeneity in the plane perpendicular to the propagation direction. The value of ${\rm Re}\, \theta(L)$ strongly depends on the frequency of the electromagnetic wave, traveling distance $L$ and on the initial polarization angle $\theta_0$.

Some obvious results that follow from equation  (\ref{finalangle}) above. It is clear, that for the particular values $v_{2d}L=\pi n$ ($n=0, 1, \dotsc$) the ${\rm Re}\, \theta(L)=\theta_0$.
In the limit $2v_{2d}L<<1$ the ${\rm Re}\, \theta(L) \approx \theta_0$ and completely independent of a particular value of $\theta_0$.
Next, if the initial angle $\theta_0 \to \pi/4$  ( $c\gg 1$- maximum difference between two orthogonally polarized eigenmodes of polarization) the ${\rm Re}\, \theta(L) \to \pi/4$ for all $L$. This means that the initial wave polarization at $\pi/4$ does not change when crossing  the medium and becomes independent of the traveling distance.

However, a significant rotation of ${\rm Re}\, \theta(L)$ from the initial value $\theta_0$ occurs when $\cos 2vL$ being positive changes the sign. In fact, if $\cot{2\theta_0} \to \infty$, then the $Re\theta(L)$ asymptotically tends to zero, if $\cos 2v_{2d}L>0$ and tends to $\pi/2$ if $\cos 2v_{2d}L<0$. Hence, for the particular value of $\theta_0=0$ (or $c=1$), the jump from $0$ to $\pi/2$ of the ${\rm Re}\, \theta(L)$ occurs at $v_{2d}L=\pi/4$. This quite drastic change of the initial polarization can be completely controlled having the appropriate parameters $L$ and frequency-dependent $v$ (see Fig. 2). In a special case of $\cos 2v_{2d}L=-1$ the $\theta(L)=\pi/2-\theta(L=0)$ and we recover the result of Ref. \cite{dielpolrot12}.

Note that capsize effect takes place for all values of $2vL$ in the range $[\pi/2, 3\pi/2]$ including the value $2vL=\pi$, considered in Ref. \cite{dielpolrot12}. When $vL=\pi/4$, as we mentioned, ${\rm Re}\, \theta(L)=\pi/4$ and independent of the initial polarization, beside $\theta_0=0$.

As another manifestation of the drastic change of the polarization, one can study the behavior of the imaginary portion of the $\theta(L=0)$ (the ellipticity or the ratio of ellipse axes) at resonant frequencies $\cos 2v_{2d}(k_0)L=0$. The latter can be written in the compact form, based on Eqs. (\ref{angle}) and (\ref{inangle})
\begin{equation}\label{im2d}
{\rm Im}\, \theta(L)  =\frac{1}{2}\ln|\tan\theta_0|.
\end{equation}
This expression provides useful information about the ratio of ellipse axes (within the sign accuracy) near the particular angle $\theta_0 \approx 0$. Exactly at $\theta_0=0$ the imaginary portion, ${\rm Im}\, \theta(L)$, tends to infinity. This means that light remains linearly polarized after traveling distance $L$ while the polarization direction rotates by $\pi/2$.

\subsection{One dimensional photonic crystal}

Knowing the explicit form of the inhomogeneity term (\ref{inhom}) it is not
difficult to calculate the splitting interaction $\hat{V}$ in the $1d$ case
\begin{equation}\label{spinorb1d}
  \hat{V}_{1d}=\frac{1}{2k_0\varepsilon}
\begin{pmatrix}
  0 & 0\\
  0 & -\frac{d\varepsilon}{d x}\frac{d}{d x}
\end{pmatrix}.
\end{equation}
The  splitting parameter $  v_{1d}$ reads
\begin{equation}
  v_{1d}=\frac{1}{2k_0}\int_{0}^{a}dx \frac{u(x)}{\varepsilon(x)}\frac{d\varepsilon}{d x}\frac{d u(x)}{d x},
\end{equation}
where $\varepsilon(x)=1+(\varepsilon-1)\Theta(x-(a-b))$ is the $1d$ periodic dielectric constant in a unit cell $[0,a]$.
Using (\ref{wavefun}) and evaluating this integral, we find
\begin{equation}\label{spl1d}
v_{1d}=\frac{A_{1d}^2\alpha(\varepsilon-1)}{2k_0\varepsilon}\tan\frac{\alpha b}{2},
\end{equation}
where $A_{1d}$ is the normalization parameter with dimensionality $a^{-1/2}$. Comparing splitting parameters in 2d, equation (\ref{split2d}), and 1d, equation (\ref{spl1d}), one can see that the 2d splitting parameter consists of an additional $b/a\ll 1$ multiplier. As a consequence, the small $v_{2d}$ in dilute photonic crystals leads to a large optical path difference between the two polarizations and make difficult to observe experimentally a capsize in 2d dilute crystals. This statement was supported by our numerical calculations of the splitting parameters $v_{1d}$ and  $v_{2d}$ versus $k_0$ using the structural parameters of the photonic crystal, used in experimental purposes. Indeed, as seen from Fig. 6, for $b=0.5$ mm, $a=10$ mm, and $\varepsilon=10$ there is a large factor in between $v_{1d}$ and $v_{2d}$ parameters. As we will see below, due to the small value of $v_{2d}$ in the frequencies range of our set up, we were not able clearly see the capsize in 2d case. However, in 1d case the capsize takes place precisely at the resonance frequencies, predicted by theory.

In a similar fashion, demonstrated in the previous section, while calculating the real part of rotation angle $Re\theta(L)$ (see equation  \ref{finalangle}), one can find PR angle for $1d$ case
\begin{eqnarray}
\label{rotangle1d}
  {\rm Re}\, \theta(L) &=& \frac{1}{2} {\rm arccot}\frac{\cot 2\theta_0}{|\cos v_{1d}L|}.
 \end{eqnarray}
The only difference between the two equations (\ref{finalangle}) and (\ref{rotangle1d}) (without taking into account the factor 2) is that the ${\cos v_{1d}L}$ in 1d case appears in the denominator. A direct consequence of this flip is that the drastic change of the polarization now takes place at $v_{1d}L= \pi/2+\pi n, n=0,\pm 1,..$. In this case ${\rm Re}\, \theta(L)=0$ if $0<\theta_0<\pi/4$ and ${\rm Re}\, \theta(L)=\pi/2$ if $\pi/4<\theta_0<\pi/2$. So for the resonant frequencies $\cos v_{1d}(k_0)L=0$, ${\rm Re}\, \theta(L)$ acquires only two values $0$ and $\pi/2$ depending on initial polarization angle $\theta_0$. Capsize takes place exactly at $\theta_0=\pi/4$.

As for the imaginary part of rotation angle that describes the ellipticity of transmitted wave at resonance frequencies $\cos v_{1d}(k_0)L=0$ one can write
\begin{equation}\label{im1d}
 {\rm Im}\, \theta(L)=\frac{1}{2}\ln\sqrt{2}|\tan(\pi/4-\theta_0)|.
\end{equation}
The infinity value of imaginary part at $\theta_0=\pi/4$ is an clear evidence of a linear polarization and a change of an initial polarization {\bf to} $\pi/2$.

\begin{figure}
 \begin{center}
\includegraphics[width=8.0cm]{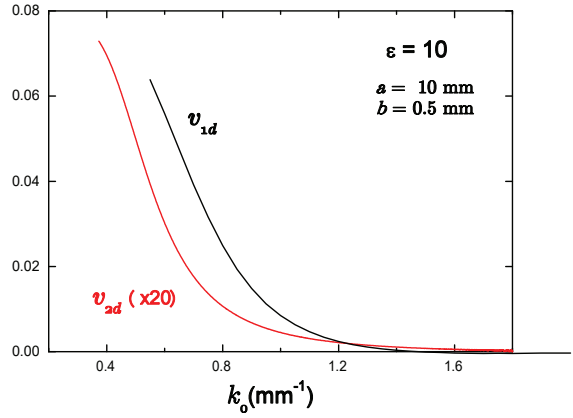}
\caption{Splitting parameter. The wavenumber dependence of the splitting parameter $v$ for one and two
dimensional crystals. In order to make a direct comparison $v_{2d}$ has been multiplied by a factor 20.}
\label{fig.6}
\end{center}
\end{figure}

\section{Conclusions}

We have carried out an experimental and theoretical investigation on the PR effects in dilute photonic crystals with transverse permittivity inhomogeneity perpendicular to the traveling direction of waves. For 1d DPC the reported experimental data show that at $\theta_0=\pi/4$ occurs capsize of an initial direction in the frequency range $10\div 140$ GHz. At the resonant frequencies $v_{1d}(k_0)L=\pi/2+\pi n$ the initial direction rotates by $\pi/4$ and coincides with $\pi/2$, in full agreement with the theoretical result (see equation  (\ref {rotangle1d})). In order to understand the capsize effect in 1d and 2d DPC a theoretical model, based on Maxwell's equations with a spatially dependent 2d inhomogeneous dielectric permittivity $\varepsilon(x,y)$, have been developed.

For one and two dimensional dilute photonic crystals we have calculated the real and imaginary parts of rotation angle. According to our calculations a capsize of initial polarization take place at resonance frequencies $v_{1d}L=\pi/2+\pi n$ and $v_{2d}L=\pi/4+\pi n$, in 1d and 2d, respectively. In the 1d case the jump of real part of rotation angle  to $\pi/2$ occurs at $\theta_0=\pi/4$ (in full agreement to our experimental data, see Fig. 3), while in 2d crystals it happens at  $\theta_0=0$. In order to give a more complete analysis of the capsize effect in these systems, we have also demonstrated that at the mentioned resonance frequencies and initial angles, the imaginary parts of the rotation angle tends to infinity. This reflect the fact that in both cases waves remains linearly polarized after traveling distance $L$ while the polarization direction rotates by $\pi/2$ and $\pi/4$ in 2d and 1d cases, respectively. The main contribution to the transmission coefficient of central diffracted wave is connected with 2d extended states that are close to the center $q = 0$ of the 2d Brillouin zone. This means  that in order to observe the above mentioned peculiarities in periodical systems, it is enough that the system exhibits long or quasi-long-range structural order in $xy-$plain \cite{GGC16}.

We have considered polarization rotation effects in dielectric DPC. However the same effects take place in metallic DPC. The only difference is the specific form of splitting parameter $v$ which should be calculated using Bloch wave functions for metallic case \cite{GGC16}. Note that optical activity in the form of circular dichroism was earlier considered in the $2D$ array of metallic spheres \cite{yanopapas09}.

\section*{Acknowledgement}

Zh.G. is grateful to A. Hakobian for his technical help and to E.Yuzbashyan and T.Shahbazyan for helpful discussions. E.C. and V.G. thank partial financial support by the Murcia Regional Agency of Science and Technology (project 19907/GERM/15). V.G. acknowledges the kind hospitality extended to him at Murcia University during his sabbatical leave.

\section*{Author contributions statement}
A.H. conducted the experiment, Zh.G. and V.G. develope the theory, E.C. carry out the numerical calculations. All the authors review the manuscript.

\section*{Additional information}
Authors have not competing financial interests.



\begin{thebibliography}{99}
\bibitem{Joan08} Joannopoulos,John D., Johnson,Steven G.,  Winn Joshua N. and  Meade, Robert D. Photonic Crystals:
Molding the Flow of Light. (Princeton University Press ,2008).
\bibitem{solli2004} Solli,D.R. and Hickmann,J.M. Photonic crystal based polarization control
devices. J. Appl. Phys. {\bf 37}, R263, (2004).
\bibitem{li2001}  Li,L.M. Two dimensional photonic crystals: candidate for wave plates. Appl.
Phys. Lett. {\bf 78}, 3400-3402, (2001).
\bibitem{solli2003} Solli,D.R., McCormick,C.F. and Chiao,R.Y. Photonic crystal polarizers and
polarizing beam splitters. J. of Appl. Phys. {\bf 93},  9429-9431, (2003).
\bibitem{shaini} Shani,Y., et al
Polarization rotation in asymmetric periodic loaded rib waveguides. Appl. Phys.
Lett. {\bf 59},  1278-1280, (1991).
\bibitem{Rytov} Rytov,S.M. On transition from wave to geometrical optics. Dokl. Akad. Nauk SSSR {\bf 18}, 263–266 (1938).
\bibitem{Vlad}Vladimirskii,V.V. The rotation of a polarization plane for curved light ray. Dokl. Akad. Nauk SSSR {\bf 21},222–225 (1941).
\bibitem{Kravor80} Kravtsov,Yu.A. and Orlov,Yu.I. Geometrical Optics of
Inhomogeneous Media. (Nauka, Moscow, 1980).
\bibitem{Vinit90} VinitskiÏ,S.I., et al., Usp.
Fiz. Nauk {\bf 160 (6)}, 1 (1990) [Sov. Phys. Usp. {\bf 33}, 403
(1990)].
\bibitem{Bliokh03}Bliokh K.Yu. and Stepanovskii, Yu.P. On the Change in Electromagnetic Wave Polarization in a Smooth One-Dimensionally Inhomogeneous Medium, JETP, {\bf 97}, 479-484, (2003).
\bibitem{dielpolrot12} Bayat and Baroughi (2012). Photonic crystal for polarization rotation, Photonic Crystals-Innovative Systems, Lasers and Waveguides, Ed.by A.Massaro.ISBN:978-953-51-0416-2,Intech,Available from; http://www.intechopen.com/books/photonic crystals.
\bibitem{metalphoton}Fumiaki Miyamaru,Kendo T., Nagashima T., and Hangyo M., Large polarization change in two-dimensional metallic photonic crystals in subterahertz region, Applied Phys.Lett., {\bf 82}, 2568, (2003).
\bibitem{GGC16}Gevorkian,Zh., Gasparian V.and Cuevas,E. Bloch states in light transport through a perforated metal. EPL, {\bf 113} 64003, (2016).
\bibitem{FF79} Feit M. D. and Fleck J. A. jr., Appl. Opt., {\bf 18},2843, (1979); Van Dyck D., Advances in Electronics and Electron Physics, {\bf 65},295, (Academic, New York, 1985) 1985.
\bibitem{Lag}De Raedt H., Lagendijk Ad and de Vries P.,Transverse Localization of Light, Phys.
Rev. Lett., {\bf 62}, 47, (1989).
\bibitem{yanopapas11} Yannopapas,V.,Photonic analog of a spin-polarized system with Rashba spin-orbit coupling, Phys.Rev.B, {\bf 83}, 113101,(2011).
\bibitem{modopt95} Karathanos,V., Stefanou N., and  Modinos A.,Optical Activity of Photonic Crystals. Journal of Mod.Opt ,{\bf 42}, 619,(1995).

\bibitem{Rashba15} Bercioux, D. and  Lueignano,P. Quantum transport in Rashba spin-orbit materials. A review,
Reports on Progress in Physics {\bf 78},106001,(2015).
\bibitem{Libzel92}Liberman,V.S. and Zel’dovich,B.Y. Spin-orbit interaction of a photon in an inhomogeneous medium. Phys.Rev. A {\bf 46}, 5199–5207 (1992).
\bibitem{Bliokh15} Bliokh,K.Y., Rodriguez-Fortu$\tilde{n}$o,F.J.,  Nori,F. and  Zayats,A.V. Spin-orbit interactions of light. Nature Photonics, {\bf 9}, 796-808, (2015).
\bibitem{yanopapas09} Yannopapas, V.,Circular dichroism in planar nonchiral plasmonic metamaterials.,Opt.Lett., {\bf 34},632,(2009).
\end{thebibliography}

\end{document}